\pdfoutput=1
\documentclass[10pt,journal,compsoc]{IEEEtran}
\usepackage[T1]{fontenc}
\usepackage{amsmath,amssymb}
\usepackage{graphicx}
\usepackage{booktabs}
\usepackage{url}
\usepackage{tikz}
\usepackage{pgfplots}
\pgfplotsset{compat=1.17}
\usetikzlibrary{positioning,arrows.meta,shapes.geometric,fit,calc}
\usepackage[hidelinks]{hyperref}
\graphicspath{{figures/}}
\newcommand{\KL}{\mathrm{KL}}

\definecolor{kb}{HTML}{E8EEF4}
\definecolor{atk}{HTML}{F8D7D4}
\definecolor{def}{HTML}{DDF0E3}
\definecolor{llm}{HTML}{FCEFC7}
\definecolor{trust}{HTML}{D8ECF6}
\definecolor{ink}{HTML}{1F3A5F}
\tikzset{
  box/.style={draw=ink, rounded corners=2pt, align=center, font=\footnotesize, inner sep=3pt, minimum height=6.5mm},
  arr/.style={-{Latex[length=1.6mm]}, draw=ink, line width=0.5pt},
  lab/.style={font=\scriptsize, align=center},
}

\makeatletter

\long\def\@makecaption#1#2{%
\ifx\@captype\@IEEEtablestring%
\footnotesize\bgroup\par\centering\@IEEEtabletopskipstrut{\normalfont\sffamily\footnotesize #1}\\{\normalfont\sffamily\footnotesize #2}\par\addvspace{0.5\baselineskip}\egroup%
\@IEEEtablecaptionsepspace
\else
\@IEEEfigurecaptionsepspace
\hbox to\hsize{\normalfont\sffamily\footnotesize #1~--~#2\hfil}%
\fi}
\makeatother

\begin{document}

\title{Detecting Word-Substitution Poisoning of a RAG Knowledge Base with Relative Entropy and Runtime Verification}

\author{Ali~Yousefi, Nikos~Komninos, and Hafizul~Asad\\[0.6em]
\normalsize Department of Computer Science, School of Science and Technology\\
\normalsize City St George's, University of London (UK)\\[0.3em]
\small \texttt{yousefi.ali@icloud.com}, \texttt{Nikos.Komninos.1@city.ac.uk}, \texttt{Hafizul.Asad@citystgeorges.ac.uk}}

\markboth{Preprint, 2026}{Yousefi \MakeLowercase{\textit{et al.}}: Detecting Word-Substitution Poisoning of a RAG Knowledge Base}

\IEEEtitleabstractindextext{%
\begin{abstract}
A retrieval-augmented generation (RAG) system answers questions from passages it retrieves from a knowledge base, so the knowledge base has to be trusted. We study an attacker who quietly edits existing passages by swapping words for synonyms (PWWS, TextFooler), in a cybersecurity RAG system built from MITRE ATT\&CK and CAPEC. Using a reviewed 44-question benchmark, repeated paired runs and a control corpus in which only the layout changes, we find that poisoning lowers judged answer correctness from 0.714 to 0.555. The damage comes mostly from lost evidence and refusals, not from false statements. To find the edited passages we use relative entropy. Comparing a whole passage with normal text is a weak signal, but one term of the relative entropy, the log-ratio between a word and its most probable synonym, works well: with a small language model as the reference it reaches an AUROC of 0.95 and stays near chance on the control. Used as a retrieval filter, it halves how often edited text reaches the generator, and an attacker who must keep each passage as probable under the language model fails on all 57 sections it tries. Filtering does not bring correctness back, because the edited passage is usually the one that held the answer. We therefore add a runtime monitor that checks each answer against trusted MITRE STIX data and remembers the passages it rejects. On its own it reduces exposure to edited text from 0.41 to 0.26; when it may add the trusted MITRE entry as a source, correctness rises to 0.786. We report the costs and failure cases of every defence alongside its gains.
\end{abstract}

}

\maketitle
\IEEEdisplaynontitleabstractindextext
\IEEEpeerreviewmaketitle

\IEEEraisesectionheading{\section{Introduction}\label{sec:intro}}
\IEEEPARstart{L}{arge} language models are now used as assistants in security work: they explain attack techniques, summarise threat reports and suggest mitigations. On their own they can be out of date or simply wrong, so most deployed assistants use retrieval-augmented generation (RAG)~\cite{lewis2020rag}. A RAG system keeps a knowledge base of trusted documents, retrieves the passages most similar to the user's question, and asks the language model to answer using only those passages. This grounds the answer in sources that an analyst can inspect.

RAG moves the trust problem rather than removing it. The answer is now only as reliable as the retrieved passages, so anyone who can change the knowledge base can change the answers. Most research on this threat studies \emph{injection}: the attacker adds new passages, written so that they are retrieved for a target question and lead the model to a chosen answer~\cite{zou2025poisonedrag,zhong2023poisoning}. Several defences have been designed for this case, for example by answering from each retrieved passage separately or by clustering the retrieved passages~\cite{xiang2024robustrag,zhou2025trustrag}.

This paper studies a quieter attacker, one who \emph{edits} passages that are already in the knowledge base. Word-level adversarial attacks such as PWWS~\cite{ren2019pwws} and TextFooler~\cite{jin2020textfooler} swap words for synonyms until a classifier changes its decision. The edited text keeps its source, its position and its topic, and no new document appears: ``implants malicious software'' becomes ``engraft malicious package''. Ikbarieh et al.~\cite{ikbarieh2025rag} showed that TextFooler edits of this kind, applied to the knowledge base of an LLM-based IoT security framework, made its attack analysis and mitigation advice worse. For a security assistant this matters: a knowledge base may be assembled from external feeds and shared repositories, and an edit inside an existing document is harder to notice than a new document.

Three things are not known about this attack. First, how much damage it does to a RAG system and \emph{how}: does the model repeat the edited words as facts, or does something else go wrong? Second, whether the edited passages can be found at retrieval time, without access to the classifier that the attacker used. Third, what a defence should do once it has found them, since removing a poisoned passage may also remove the evidence the answer needed.

We address these questions with AegisRAG, a cybersecurity question-answering system built from MITRE ATT\&CK and CAPEC, in which every experiment is recorded and can be re-checked. Our detector is based on relative entropy. If a passage's word distribution $p$ is compared with a reference $q$ for normal text, $\KL(p\,\|\,q)=\sum_x p(x)\log_2\frac{p(x)}{q(x)}$~\cite{kullback1951} should rise when words are swapped for unusual synonyms. In our data the full sum is a weak signal, because only a few of the roughly eighty words in a passage change. One term of the sum is much stronger: the log-ratio between a word and its most probable synonym under $q$. We call it the \emph{synonym gap}. It builds on the observation behind frequency-guided substitution detection~\cite{mozes2021fgws}, that adversarial substitutes tend to be rarer than the words they replace, but it needs only the text and a reference, so it can run inside the retrieval step.

Detection alone turns out not to be enough. When we remove the passages the detector flags, less poisoned text reaches the model, but the answers do not get better, because the model is left without evidence and declines to answer. We therefore also test a different kind of defence, a \emph{runtime monitor} in the sense of runtime verification~\cite{leucker2009rv,bartocci2018rv}. It checks each answer against properties derived from the official MITRE data, which the attacker cannot edit, sends violations back to the model, and keeps a memory of the passages it has stopped trusting.

Our main findings are as follows.
\begin{itemize}
\item \textbf{The attack removes evidence more than it adds lies.} PWWS poisoning lowers judged correctness from 0.714 to 0.555, almost entirely on questions whose source section was edited. Refusals rise from 26 to 45 of 132 answers, while wrong answers rise only from 0 to 2 (Section~\ref{sec:mechanism}).
\item \textbf{A local relative-entropy score finds the edits.} The synonym gap with a small language model as reference reaches an AUROC of 0.95 on PWWS and 0.91 on TextFooler, and stays near chance (0.53) on a layout-only control. It is clearly better than whole-passage perplexity, similar to a windowed perplexity filter, and it names the word it suspects (Section~\ref{sec:detection}).
\item \textbf{Filtering makes the system safer but not more useful.} Dropping flagged passages halves exposure to edited text and removes the wrong answers, but correctness does not recover (Section~\ref{sec:defences}).
\item \textbf{Attackers who know the detector struggle.} An attacker that must keep each passage as probable under the language model succeeds on none of 57 sections (Section~\ref{sec:adaptive}).
\item \textbf{A trusted source, not detection, restores answers.} The runtime monitor reduces exposure to edited text from 0.41 to 0.26, and with a trusted MITRE fallback correctness rises to 0.786. Its memory never re-admits a quarantined passage, but its blame step also quarantines some clean passages (Section~\ref{sec:monitor}).
\end{itemize}

The rest of the paper is organised as follows. Section~\ref{sec:related} reviews related work. Section~\ref{sec:system} describes the system, the attacker and the benchmark. Section~\ref{sec:method} presents the detectors, the retrieval filters and the runtime monitor. Section~\ref{sec:eval} reports the experiments. Section~\ref{sec:discussion} discusses what the results mean, their limitations and ethics, and Section~\ref{sec:conclusion} concludes.

\section{Background and Related Work}\label{sec:related}
\textbf{Poisoning RAG.} PoisonedRAG~\cite{zou2025poisonedrag} injects passages that are retrieved for a target question and lead to a chosen answer. Zhong et al.~\cite{zhong2023poisoning} craft passages that are retrieved for many queries. Both add text. Ikbarieh et al.~\cite{ikbarieh2025rag} instead edit an existing knowledge base with TextFooler. We study the same kind of in-place edit and measure where, between retrieval and generation, the damage happens.

\textbf{Word-substitution attacks and their detection.} PWWS~\cite{ren2019pwws} ranks words by how much they matter to a classifier and replaces each with the WordNet~\cite{miller1995wordnet} synonym that most lowers the classifier's confidence. TextFooler~\cite{jin2020textfooler} takes substitutes from counter-fitted word embeddings. Both are available in TextAttack~\cite{morris2020textattack}. FGWS~\cite{mozes2021fgws} detects such examples by replacing infrequent words with frequent synonyms and checking whether the classifier's output changes. We keep its core observation but score the text against a reference distribution instead of querying the attacked classifier.

\textbf{Perplexity filters.} Language-model perplexity is a common, model-agnostic filter for adversarial text. Adversarial suffixes have very high perplexity~\cite{alon2023perplexity}, and windowed perplexity is a strong baseline against them~\cite{jain2023baseline}. RAGuard~\cite{cheng2025raguard} widens retrieval and filters passages by chunk-wise perplexity. We compare against whole-passage and windowed perplexity computed with the same language model as our detector.

\textbf{Defending RAG.} RobustRAG~\cite{xiang2024robustrag} answers from each passage separately and aggregates the answers, with certifiable guarantees. TrustRAG~\cite{zhou2025trustrag} clusters retrieved passages to find suspicious groups. Corrective RAG~\cite{yan2024crag} grades retrieval and falls back to other sources when it is poor. RAGAS faithfulness~\cite{es2024ragas} checks whether an answer follows from its sources, and SelfCheckGPT~\cite{manakul2023selfcheckgpt} uses sampling to find hallucinations. These defences are mainly aimed at injected passages or at answer quality. We filter \emph{edited} passages before generation and check answers against a trusted source afterwards.

\textbf{Runtime verification.} Runtime verification checks a running system against formally stated properties, using a monitor that observes its behaviour~\cite{leucker2009rv,bartocci2018rv}. Properties over time are commonly written in linear temporal logic (LTL)~\cite{pnueli1977ltl}. We apply the idea to the output of a RAG system: the monitor observes each draft answer and its context, and one of its properties is a temporal safety property about which passages may be used.

\section{System, Threat Model and Benchmark}\label{sec:system}

\begin{figure*}[t]
\centering
\resizebox{0.98\textwidth}{!}{%
\includegraphics{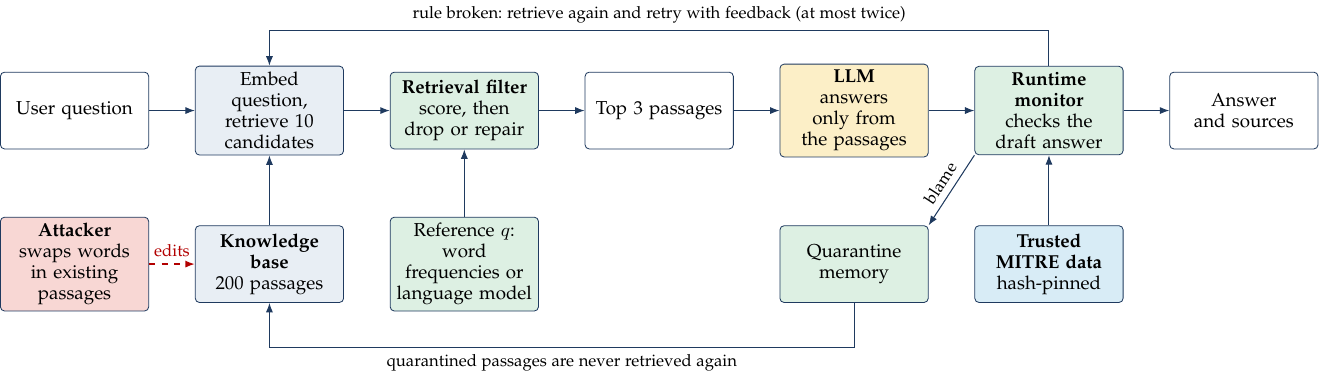}}
\caption{AegisRAG overview}\label{fig:overview}
\end{figure*}

\subsection{The RAG system}
Figure~\ref{fig:overview} shows the system. A FastAPI backend embeds the question with all-MiniLM-L6-v2, retrieves passages from a Qdrant vector store, and asks gpt-4o-mini (temperature 0) to answer, using a fixed prompt that tells the model to answer only from the context and to say so when the context is not enough. Without a defence, the 3 most similar passages go to the model. Documents are split into 500-character passages with 100 characters of overlap. A web interface shows answers from the clean and the poisoned knowledge base side by side and lets the user switch each defence on.

The knowledge base contains 18 documents of MITRE ATT\&CK and CAPEC material on 14 attack types in 5 categories (DDoS, injection, malware, reconnaissance and spoofing), split into 200 passages.

\subsection{Threat model}
The attacker can edit text in the knowledge base. It cannot change the questions, the retrieval or generation settings, the trusted MITRE files or the evaluation. Its goal is to make the system's answers worse while keeping the edits hard to notice.

\begin{figure}[t]
\centering
\resizebox{\columnwidth}{!}{%
\includegraphics{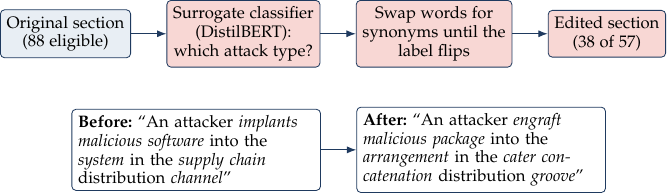}}
\caption{How the knowledge base is poisoned}\label{fig:attack}
\end{figure}

Figure~\ref{fig:attack} shows how we build the poisoned knowledge base. We train a DistilBERT classifier to predict the attack type of a section, on the same material. It acts as a \emph{surrogate}: the attacker does not attack the RAG system directly but rewrites a section whenever PWWS (run with TextAttack) flips the classifier's label. Of 88 eligible sections, 31 are already misclassified and are skipped, and PWWS succeeds on 38 of the remaining 57. The adaptive experiments in Section~\ref{sec:adaptive} use a retrained surrogate, on which plain PWWS succeeds on 39. The original and the edited text of every section are kept in an attack manifest.

\textbf{Layout-only control.} The poisoning script also flattens the layout of the sections it rewrites. To separate the effect of the words from the effect of the layout, a control corpus applies the same flattening to the same sections but keeps the original words. Retrieval on the control is the same as on the clean corpus, so the layout change alone does not explain the damage.

\subsection{Benchmark and protocol}
The benchmark has 44 reviewed questions. Each has a list of reference facts and the exact source section that answers it. Development and held-out questions come from different source documents (35 and 9 questions).

Each question is run 3 times in every condition. Runs are paired: the same question is asked of two systems in the same schedule slot, with the order alternated. We measure:
\begin{itemize}
\item \emph{Correctness} (0 to 1): an LLM judge~\cite{zheng2023judge} compares the answer with the reference facts using a fixed rubric. On 75 answers that the first author graded blind, the judge agreed well with the manual grades (quadratic weighted $\kappa=0.82$).
\item \emph{Hit@3}: whether a passage from the right section is among the top 3.
\item \emph{Exposure}: whether any attacker-edited word reached the language model.
\item \emph{Refusals} and \emph{wrong answers}: an answer is a refusal if it says the context is insufficient, and wrong if it is not a refusal and its correctness is below 0.25.
\end{itemize}
Retrieval returned the same passages in all three repetitions of every question, so the question is the unit of analysis. We report 95\% confidence intervals from a question-level bootstrap (10{,}000 resamples) and two-sided Wilcoxon signed-rank tests. Every run writes one JSON record per answer, which is never overwritten, and a completion file with the SHA-256 hash of every record.

\section{Defences}\label{sec:method}
\subsection{Relative-entropy detectors}\label{sec:detectors}
Let a passage contain the words $w_1,\dots,w_n$ with empirical distribution $p$, and let $q$ be a reference distribution for normal text.

\textbf{Whole-passage scores.} As baselines we compute $\KL(p\,\|\,q)$~\cite{kullback1951}, the Jensen--Shannon divergence~\cite{lin1991js} and the mean surprisal $-\frac{1}{n}\sum_i \log_2 q(w_i)$. We also report KL over letter frequencies, the usual textbook example of relative entropy.

\begin{figure}[t]
\centering
\resizebox{\columnwidth}{!}{%
\includegraphics{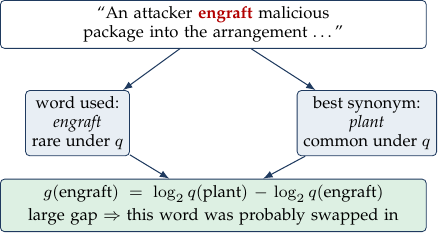}}
\caption{The synonym gap for one word}\label{fig:gap}
\end{figure}

\textbf{Synonym gap.} For each content word $w$ with WordNet synonyms $S(w)$ we compute
\begin{equation}
g(w)=\max_{s\in S(w)}\big[\log_2 q(s)-\log_2 q(w)\big],
\end{equation}
the number of bits by which the most probable synonym beats the word that was used (Figure~\ref{fig:gap}). This is the log-ratio term of KL, evaluated where a substitution would have happened. A passage's score is the mean of its three largest gaps.

\textbf{References.} The score depends on $q$, and we compare four choices. (i) General English word frequencies. (ii) A \emph{domain} reference: word counts from the rest of the knowledge base, leaving out the passage's own document, mixed 50/50 with English. It does not need a clean copy of the poisoned document. (iii) The same counts with a Dirichlet prior, $q(w)=\frac{c(w)+\mu q_{\mathrm{en}}(w)}{N+\mu}$ with $\mu=2000$~\cite{zhai2004smoothing}. (iv) A \emph{contextual} reference from a small language model, distilgpt2~\cite{radford2019gpt2}, which judges a word in its sentence rather than on its own. For (iv) we take the 10 words with the largest domain gap and, for each, its 5 most probable synonyms $s$, and compute
\begin{equation}
g_{\mathrm{ctx}}(w,s)=\log_2 p_{\mathrm{LM}}(\text{passage}, w\!\to\! s)-\log_2 p_{\mathrm{LM}}(\text{passage}),
\end{equation}
that is, how much more probable the whole passage becomes when $w$ is replaced by $s$. The passage score is again the mean of the three largest gaps.

\textbf{Perplexity baselines.} With the same language model we compute the \emph{mean surprisal} (log$_2$ perplexity of the whole passage) and the \emph{top-5 surprisal}, the mean of the five most surprising tokens, which is a windowed filter in the spirit of~\cite{jain2023baseline}.

All scores use only information a defender would have. The attack labels are used only for evaluation. Thresholds were set once, on development documents, at a 10\% false-positive rate, and were not changed afterwards.

\subsection{Retrieval filters}\label{sec:filters}
A filter retrieves 10 candidates instead of 3, scores them, and passes 3 on to the language model (Figure~\ref{fig:overview}). We test three:
\begin{itemize}
\item \emph{Drop (frequency)} removes candidates whose word-frequency synonym gap is above the threshold (7.23 bits).
\item \emph{Repair} first replaces each word whose gap is above 9.5 bits with its likely original, and keeps the passage if it then scores under the threshold.
\item \emph{Drop (contextual)} removes candidates whose contextual synonym gap is above its threshold (4.46 bits).
\end{itemize}

\subsection{Runtime monitor}\label{sec:monitor-design}
The filters have no memory: the same poisoned passage is scored again each time it is retrieved. The runtime monitor checks each draft answer, removes the passages it blames, and remembers them. Its trusted reference is the official MITRE STIX data for CAPEC and ATT\&CK. The files are pinned by their SHA-256 hash, and the monitor refuses to start if they change.

\textbf{Properties.} Each draft answer is checked against three properties.
\begin{itemize}
\item $\varphi_1$ (\emph{no unknown rare words}): every rare word in the answer (Zipf frequency below 3.5) that was copied from the retrieved passages occurs in the trusted MITRE vocabulary or in the question.
\item $\varphi_2$ (\emph{no unnecessary refusal}): the answer is not a refusal when the trusted reference covers the attack being asked about.
\item $\varphi_3$ (\emph{never trust again}): $\mathbf{G}\big(\mathit{quarantined}(p)\rightarrow\mathbf{G}\,\neg\mathit{inContext}(p)\big)$. Once a passage is quarantined it never appears in the context again.
\end{itemize}
$\varphi_1$ and $\varphi_2$ are checked by code on each answer. $\varphi_3$ is a temporal safety property written in LTL; it holds by construction, because quarantined passages are skipped at retrieval, and we also check it on every recorded trace.

\begin{figure}[t]
\centering
\resizebox{\columnwidth}{!}{%
\includegraphics{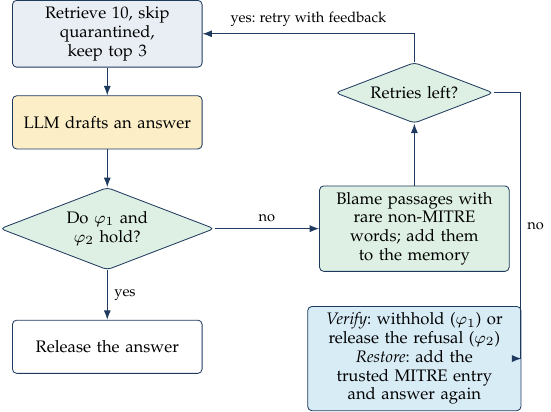}}
\caption{The runtime monitor loop}\label{fig:monitor}
\end{figure}

\textbf{Loop.} Figure~\ref{fig:monitor} shows the loop. When $\varphi_1$ or $\varphi_2$ is violated, the monitor blames the retrieved passages that contain rare words outside the trusted vocabulary and adds them to an append-only memory, keyed by a hash of each passage's source and text. It then retrieves again without them and asks the model again, giving it the rejected answer and the violated property as feedback. This is repeated at most twice.

\textbf{Modes.} In \emph{verify} mode, a remaining $\varphi_1$ violation withholds the answer and a remaining $\varphi_2$ violation releases the refusal. In \emph{restore} mode, a remaining $\varphi_2$ violation instead adds the trusted MITRE entry for the attack as an extra source and the model answers again. Restore mode gives the model information that the clean system never sees, so we treat it as an upper bound on what a trusted source can add, not as a defence comparable with the filters.

\section{Evaluation}\label{sec:eval}
We evaluate in the order of the questions in Section~\ref{sec:intro}: the attack, detection, the filters, attackers who know the detector, a second attack, and the monitor. All live runs use the full benchmark (44 questions $\times$ 3 repetitions, 132 answers per condition).

\subsection{What the attack does}\label{sec:mechanism}
\begin{table}[t]
\caption{Effect of PWWS poisoning (44 questions, 132 paired answers per condition). $\Delta$: change in correctness.}\label{tab:attack}
\centering\footnotesize\setlength{\tabcolsep}{3pt}
\begin{tabular}{@{}lccc@{}}
\toprule
Questions & Hit@3 & Correctness & $\Delta$ [95\% CI]\\
\midrule
All (44) & .705$\to$.568 & .714$\to$.555 & $-$.159 [$-$.254, $-$.068]\\
Source section edited (27) & .630$\to$.444 & .682$\to$.432 & $-$.250 [$-$.383, $-$.124]\\
Not edited (17) & .824$\to$.765 & .765$\to$.750 & $-$.015 [$-$.083, .069]\\
\bottomrule
\end{tabular}
\end{table}

\begin{figure}[t]
\centering
\includegraphics{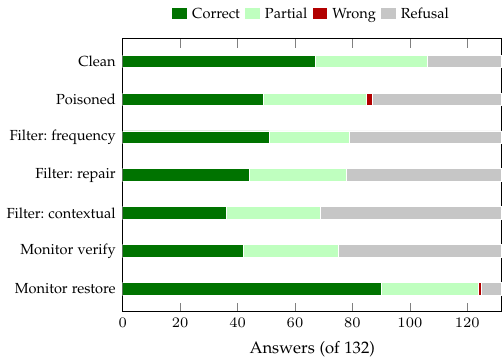}
\caption{What happens to the answers}\label{fig:outcomes}
\end{figure}

Poisoning lowers correctness from 0.714 to 0.555 (Table~\ref{tab:attack}). The drop comes from the 27 questions whose source section was edited. On the 17 questions answered from untouched sections of the same poisoned corpus, the change is small and not significant. Together with the layout-only control, this points to the word substitutions as the cause.

Most of the damage is lost evidence (Figure~\ref{fig:outcomes}). An edited passage looks less like the question, so it can fall out of the top 3, and the model, told to answer only from the context, declines. Refusals rise from 26 to 45 of 132 answers. Of the 48 answers whose correctness falls by at least 0.25, 27 are refusals and 31 have no relevant passage in the top 3. Wrong answers rise only from 0 to 2. Both are to the same question, and in both the model repeats a substituted phrase, ``Engraft Malicious Package'' (originally ``implants malicious software''), as if it were the name of an attack stage.

\subsection{Finding the edited passages}\label{sec:detection}
\begin{table*}[t]
\caption{Detection of passages that contain substituted words (AUROC: 1.0 is perfect, 0.5 is chance). Brackets: 95\% bootstrap intervals that resample documents. 191 passages (193 for TextFooler); edited passages: 47 PWWS, 59 PWWS on the retrained surrogate, 82 TextFooler, 13 frequency-matched adaptive. The control has no word edits: its ``positives'' are the 48 passages whose layout was flattened, so a score above 0.5 there reflects layout or section choice, not substitutions. Last row: edits flagged at the contextual threshold (4.46 bits) fixed on PWWS development documents, which flags 9--14\% of unedited passages.}\label{tab:detect}
\centering\footnotesize\setlength{\tabcolsep}{4pt}
\begin{tabular}{@{}lccccc@{}}
\toprule
Detector & PWWS & PWWS (retrained) & TextFooler & Control (layout only) & Frequency-matched adaptive\\
\midrule
Letter-frequency KL & 0.33 & 0.31 & 0.28 & 0.35 & 0.36\\
Word KL vs domain & 0.72 & 0.74 & 0.83 & 0.62 & 0.50\\
Jensen--Shannon vs domain & 0.47 & 0.47 & 0.48 & 0.43 & 0.44\\
Mean surprisal (perplexity) & 0.77 & 0.77 & 0.82 & 0.45 & 0.73\\
Top-5 surprisal (windowed perplexity) & 0.91 & 0.93 & 0.92 & 0.56 & 0.85\\
Synonym gap, domain $q$ & 0.82 [.72, .90] & 0.88 [.82, .94] & 0.90 [.85, .94] & 0.67 [.57, .77] & 0.47 [.33, .72]\\
Synonym gap, Dirichlet $q$ & 0.83 & 0.89 & 0.90 & 0.68 & 0.48\\
Contextual synonym gap & 0.95 [.87, .99] & 0.97 [.93, 1.0] & 0.91 [.84, .97] & 0.53 [.45, .60] & 0.92 [.54, .96]\\
\midrule
\emph{Edits flagged, fixed contextual threshold} & 42/47 & 54/59 & 70/82 & -- & 10/13\\
\bottomrule
\end{tabular}
\end{table*}

\begin{figure}[t]
\centering
\includegraphics{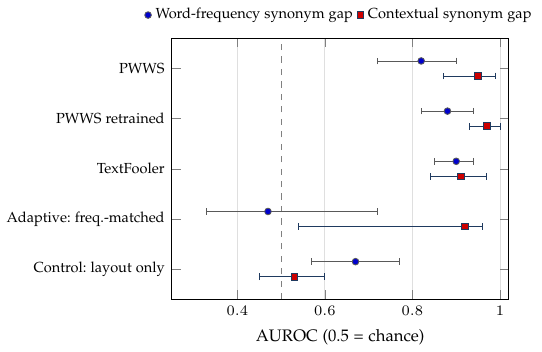}
\caption{The two synonym-gap detectors on every corpus}\label{fig:robust}
\end{figure}

We score every passage and ask how well each score separates edited from unedited passages, using the area under the ROC curve (AUROC). Table~\ref{tab:detect} and Figure~\ref{fig:robust} show four things.

\emph{Whole-passage scores are weak.} Letter frequencies keep their English statistics when words are swapped, so letter KL is below chance. Word KL against the domain reference does better (0.72 to 0.83), but it also scores 0.62 on the control, where no word changed. Part of what it picks up is therefore which sections were attacked, not the edits themselves.

\emph{The per-word log-ratio works, and the reference matters more than the formula.} The synonym gap with the domain reference reaches 0.82 on PWWS. Changing the smoothing makes almost no difference (0.83).

\emph{The language-model reference is better on PWWS.} It improves on the word-frequency gap on both PWWS corpora (paired differences $+0.13$ [$+0.03$, $+0.21$] and $+0.09$ [$+0.03$, $+0.16$]) and is close to chance on the control, where the word-frequency gap still scores 0.67. It is not better everywhere. On TextFooler the two are not distinguishable ($+0.02$ [$-0.05$, $+0.07$]), and on the 9 held-out documents (45 passages, 10 of them edited) the word-frequency gap scores higher (0.94 against 0.88).

\emph{What helps is looking locally.} The contextual gap beats whole-passage perplexity on PWWS and TextFooler (paired differences $+0.18$ [$+0.11$, $+0.26$] and $+0.10$ [$+0.03$, $+0.16$]). It is statistically indistinguishable from windowed perplexity ($+0.04$ [$-0.003$, $+0.08$] on PWWS, $-0.01$ [$-0.07$, $+0.03$] on TextFooler). What the synonym gap adds is an explanation. Of the 27 edited passages that the word-frequency gap flags, 22 name the exact replaced word and its likely original (for example abnegate$\to$deny and meshing$\to$network). At the fixed thresholds the contextual gap also has a slightly lower false-positive rate than windowed perplexity on every attacked corpus (for example 10.4\% against 11.8\% on PWWS).

\subsection{Filtering at retrieval time}\label{sec:defences}
\begin{figure}[t]
\centering
\includegraphics{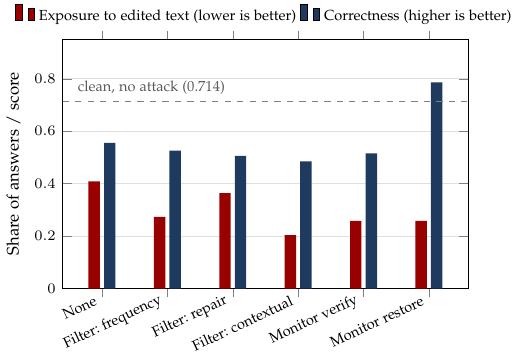}
\caption{Every defence on the PWWS-poisoned knowledge base}\label{fig:defences}
\end{figure}

\begin{table}[t]
\caption{Retrieval filters on the poisoned corpus (44 questions $\times$ 3; $p$: Wilcoxon against no defence). Clean cost: change in correctness on the clean corpus.}\label{tab:defence}
\centering\small
\setlength{\tabcolsep}{3pt}
\begin{tabular}{@{}lccccc@{}}
\toprule
Defence & Exposure & Correct. & Wrong & Refusals & Clean cost\\
\midrule
None & .409 & .555 & 2 & 45 & --\\
Drop (freq.) & .273$^{a}$ & .526 & 0 & 53 & $-$.064\\
Repair & .364 & .506 & 0 & 54 & $-$.042\\
Drop (ctx.) & .204$^{b}$ & .485$^{c}$ & 0 & 63 & $-$.013\\
\bottomrule
\multicolumn{6}{@{}l}{\footnotesize $^{a}p=0.031$; $^{b}p=0.012$; $^{c}p=0.020$.}
\end{tabular}
\end{table}

All three filters remove the two wrong answers (Table~\ref{tab:defence}, Figure~\ref{fig:defences}). The contextual filter halves how often edited text reaches the model and has the smallest cost on clean data; it flags 8.4\% of clean candidates, against 10.9\% for the word-frequency filter. None of the filters restores correctness, and the contextual filter lowers it further. The reason is visible in Figure~\ref{fig:outcomes}: when the passage that answers a question is the edited one, removing it leaves the model without evidence, and it refuses. Repair does not help either, because word-level WordNet repair restores too few substitutions (recall 0.12 at precision 0.89). Filtering makes the poisoned knowledge base safer to read, but not useful again.

We also tried checking the \emph{answer} instead of the passages: a separate verifier call splits each draft answer into claims, judges each claim against the retrieved passages, and must quote its evidence word for word. Replayed on 352 recorded answers, it was not a useful error detector. Only 28\% of the answers it flagged were wrong (precision 0.28 [0.13, 0.48]); it flagged 20\% of the incorrect answers and 31\% of the correct ones. The typical error here is an incomplete answer in which every statement is grounded (68 of 89 partial answers passed), and a grounding check cannot see what is missing.

\subsection{Attackers who know the detector}\label{sec:adaptive}
\begin{table}[t]
\caption{Adaptive attackers (same retrained surrogate and 57 attackable sections; only the evasion rule differs). AUROC of the word-frequency and contextual synonym gaps.}\label{tab:adaptive}
\centering\small
\setlength{\tabcolsep}{3pt}
\begin{tabular}{@{}lcccc@{}}
\toprule
Attacker & Sections & Edits & Freq. & Ctx.\\
\midrule
Unrestricted PWWS & 39 & 59 & .88 & .97\\
1st order: frequency-matched & 8 & 13 & .47 & .92\\
2nd order, strict & 0 & 0 & -- & --\\
2nd order, $\le$4.46 bits & 0 & 0 & -- & --\\
2nd order, $\le$20 bits & 0 & 0 & -- & --\\
2nd order, $\le$100 bits & 5 & 6 & .51 & .78\\
2nd order, $\le$300 bits & 8 & 13 & .47 & .92\\
\bottomrule
\end{tabular}
\end{table}

A detector that is published can be attacked. We test two kinds of attacker that know it (Table~\ref{tab:adaptive}).

A \emph{first-order} attacker only swaps to words that are at least as frequent under the defender's $q$. It evades the word-frequency detector, as intended. The contextual detector still finds most of its edits, because the attacker matched how common its words are, not whether they fit the sentence. Evasion also costs the attacker: it succeeds on 8 sections instead of 39.

A \emph{second-order} attacker must also keep each passage at least as probable under the detector's language model. It succeeds on no section. To see why, we looked at the 550 swaps that unrestricted PWWS used in its 39 successful attacks. 109 of them (20\%) satisfy the frequency rule, only 9 (1.6\%) leave the passage at least as probable under the language model, and 5 satisfy both. The median swap makes its passage 17.9 bits less probable. Our reading is that natural text sits close to a local maximum of language-model likelihood, and the swaps that move a classifier tend to move the text away from it.

\begin{figure}[t]
\centering
\includegraphics{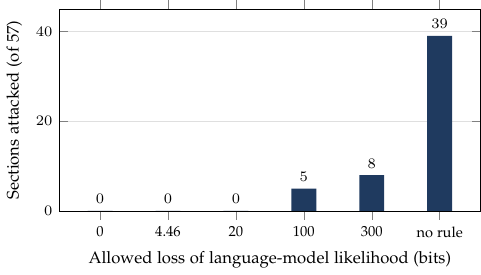}
\caption{Second-order attacker and the language-model budget}\label{fig:budget}
\end{figure}

We then relaxed the rule (Figure~\ref{fig:budget}). With an allowed loss of up to 4.46 bits (the contextual threshold) or 20 bits, the attacker still succeeds on no section. At 100 bits it succeeds on 5 sections; the contextual detector scores their 6 edited passages at an AUROC of 0.78, and the fixed filter flags 3 of the 6, so 3 edited passages get through. At 300 bits the rule no longer constrains the attacker, and it produces exactly the same corpus as the first-order attacker, whose edits the contextual filter flags 10 of 13. In this family of attackers, those that respect the detector's rule do not succeed, and those that succeed leave most, but not all, of their edits detectable. These conclusions rest on 6 and 13 edited passages.

\subsection{A second attack: TextFooler}\label{sec:textfooler}
TextFooler takes substitutes from word embeddings instead of WordNet. With the same surrogate and sections it is the stronger attack: 49 of 57 sections and 82 edited passages. Both synonym-gap detectors work on it without changes (Table~\ref{tab:detect}). At the thresholds fixed on PWWS development documents, the contextual filter flags 70 of 82 edits (85\%) at a 9\% false-positive rate, and the word-frequency filter 43 of 82 (52\%) at 6\%.

Run live with the same thresholds, TextFooler also reaches the generator more often. Edited text is in the context for 82\% of answers (PWWS: 41\%), and correctness falls from 0.714 to 0.564 ($p=0.002$), again through refusals and not wrong answers. The word-frequency filter barely changes exposure (82\% to 80\%): when most candidates are edited, dropping some of them mainly promotes others. The contextual filter cuts exposure to 52\% ($p<0.001$), at the same cost as on PWWS: correctness falls further, to 0.447, and refusals rise from 54 to 74 of 132.

\subsection{The runtime monitor}\label{sec:monitor}
We ran the monitor in four conditions: PWWS with verify mode, PWWS with restore mode, clean data with verify mode, and TextFooler with verify mode. Each condition starts with an empty memory, so repetitions 2 and 3 meet the same attack again and show what the memory adds.

\textbf{Two versions.} The first version of $\varphi_1$ flagged any rare word outside the trusted vocabulary, including words the model chose itself. In a first live run it withheld 10 answers, one of them on clean data, partly because of words from the knowledge base's OWASP and NIST material, which the MITRE files do not contain. We changed $\varphi_1$ to count only words copied from the retrieved passages and repeated the full run. Table~\ref{tab:monitor} reports the second run. Both runs are kept.

\begin{table}[t]
\caption{Runtime monitor, live, second version (44 questions $\times$ 3; question-level paired comparisons; $p$: Wilcoxon). One TextFooler answer was lost to a network error and is excluded.}\label{tab:monitor}
\centering\footnotesize\setlength{\tabcolsep}{1.5pt}
\begin{tabular}{@{}lccc@{}}
\toprule
Comparison & Monitor & Baseline & $\Delta$ [95\% CI]\\
\midrule
Verify, PWWS: exposure & .258 & .409 & $-$.152 [$-$.265, $-$.061]$^{a}$\\
Verify, PWWS: correctness & .515 & .555 & $-$.040 [$-$.085, .000]\\
Verify vs ctx.\ filter: exposure & .258 & .204 & +.053 [$-$.030, .144]\\
Verify vs ctx.\ filter: correctness & .515 & .485 & +.030 [$-$.002, .068]\\
Verify, clean data: correctness & .708 & .714 & $-$.006 [$-$.023, .011]\\
Verify, TextFooler: exposure & .583 & .818 & $-$.235 [$-$.356, $-$.114]$^{b}$\\
Verify, TextFooler: correctness & .542 & .564 & $-$.023 [$-$.087, .047]\\
Restore, PWWS: correctness & .786 & .555 & +.231 [.136, .335]$^{c}$\\
Restore vs clean: correctness & .786 & .714 & +.072 [$-$.004, .152]\\
\bottomrule
\multicolumn{4}{@{}l}{\footnotesize $^{a}p=0.016$; $^{b}p=0.001$; $^{c}p=0.0001$.}
\end{tabular}
\end{table}

\textbf{Verify mode.} The monitor reduces exposure to edited text on both attacks (Table~\ref{tab:monitor}) and leaves no wrong answers on PWWS. It does not restore correctness, which is slightly lower than without the monitor, and refusals rise from 45 to 57. Against the contextual filter it is not significantly different in exposure or correctness. On clean data it costs almost nothing, quarantines no passage and withholds no answer, although one clean answer is judged wrong where the unmonitored system had none.

The phrase ``Engraft Malicious Package'' reaches the user in 3 of 3 repetitions without a defence and in none with the monitor. In the first repetition the edited passage is retrieved, but the first draft does not repeat the phrase; the monitor quarantines that passage, and in repetitions 2 and 3 the memory keeps it out of the context. In all three repetitions the final answer is a refusal and not a correct answer, and the plain filters also removed the phrase.

\textbf{Memory.} $\varphi_3$ holds on every recorded trace in all four conditions. Quarantined passages are skipped on later questions: the number skipped at the first retrieval rises from 0.73 to 1.38 per answer between repetition 1 and repetitions 2--3 on PWWS (1.25 to 2.81 on TextFooler), and on TextFooler the number of generations per answer falls from 1.32 to 1.01. The effect on exposure is small (0.27 to 0.25 on PWWS, 0.60 to 0.57 on TextFooler), and correctness is lower in the later repetitions (0.53 to 0.51 and 0.59 to 0.52), because more passages have been removed.

The memory is only as accurate as the blame step. On TextFooler 26 of the 27 quarantined passages contain edited text, but on PWWS only 10 of 18 do. In the PWWS verify condition every quarantine was triggered by $\varphi_2$, that is, by a refusal, and the passages blamed were those with rare words outside the MITRE vocabulary. An ordinary word such as ``ingested'' was enough to quarantine an unedited passage.

\textbf{Restore mode.} With the trusted fallback, correctness rises from 0.555 to 0.786 and refusals fall from 45 to 7 (Figure~\ref{fig:outcomes}). The monitor falls back in 58 of 132 answers. The result is numerically above the unattacked system (0.714), but the difference is not significant. The gain comes from the trusted MITRE entry, which covers most of our questions because the knowledge base was built from MITRE material; it would be smaller for content without a trusted counterpart. The monitor's part is deciding when to fall back and which passages to stop using.

\section{Discussion}\label{sec:discussion}
\textbf{Why one term works better than the sum.} $\KL(p\,\|\,q)$ adds up log-ratio terms weighted by $p$. A substitution changes only a few terms, each with a small weight, and the passage's topic already makes $p$ differ from $q$ by much more. The synonym gap looks at the log-ratio only where an attack would act, comparing each word with the word it may have replaced.

\textbf{Why the reference matters most.} Changing the smoothing or using a symmetric divergence makes little difference, while moving from word frequencies to a language model makes a large one on PWWS. A word-frequency reference can only say that a word is rare. A contextual one can say that it is unlikely in this sentence. The perplexity baselines point the same way: what matters is measuring improbability locally, where the edit is.

\textbf{Detection is not restoration.} In this corpus the passage that answers a question about an edited section is usually the edited passage, so a filter that removes poison also removes evidence. Getting correctness back needs another source of the same information. The monitor's restore mode uses the trusted MITRE entry. Where no trusted source exists, the original text would have to be recovered, for example with a masked language model adapted to the domain.

\textbf{What the monitor adds.} The monitor does not detect better than the contextual filter. Its contributions are a memory with a property that can be checked, a rule for when to fall back to trusted data, and a record of why each passage was rejected. Its weakest part is blame: a refusal says that something is missing, not which passage is at fault, and blaming by rare words quarantined 8 clean passages on PWWS. Using the contextual synonym gap for blame is the natural next step.

\textbf{Limitations.} The study uses one small knowledge base (18 documents), one generation model, one embedding model and 44 questions, of which only 9 are held out. The adaptive results rest on 13 and 6 edited passages, and the confidence intervals are wide. The second-order attacker constrains whole-passage likelihood but does not optimise against the detector's score directly; such an attacker, a masked-language-model attack, or a larger reference model could behave differently. The attack targets a surrogate classifier and not the RAG pipeline itself. Correctness is scored by an LLM judge. On a blind manual check of 75 answers by the first author it agreed well with the manual grades ($\kappa=0.82$, Spearman $\rho=0.85$) but was more lenient by 0.08 on average: 0.02 on clean answers, 0.08 on poisoned answers and 0.14 on answers produced with the word-frequency filter. A second, independent grader would make this check stronger. For the monitor, $\varphi_1$ can be evaded by an attacker who chooses substitutes inside the trusted vocabulary or common enough to pass the rarity cut-off; the trusted reference covers only MITRE; and an attacker could try to get clean passages quarantined through the blame step. $\varphi_1$ and $\varphi_2$ are implemented as code and are not generated from a formal specification.

\textbf{Ethics.} The attacks edit a local copy of public MITRE material. No real system is attacked and no malware is run. No personal data is used; question and passage text is sent to a commercial LLM API. Detection scores are evidence for a person to review, not proof that a passage is clean or that an answer is true. We evaluate attackers who know the detector and report where it fails, so that it is not over-trusted.

\textbf{Reproducibility.} Code, corpora, attack manifests, run records and analysis scripts are at \url{https://github.com/Aliyousefi9403/dissertation-RAG/tree/dissertation-v1.0} (branch \texttt{dissertation-v1.0}, the version this paper describes). A verification script re-hashes all run records and checks the main numbers reported here against the result files that produced them.

\section{Conclusion}\label{sec:conclusion}
We studied an attacker who poisons a cybersecurity RAG knowledge base by swapping words for synonyms. The attack does most of its damage by removing evidence: the model refuses more often and is wrong only occasionally. A local relative-entropy score, the synonym gap, finds the edited passages, most reliably with a small language model as its reference, and attackers constrained to evade it lose most or all of their successful attacks. Filtering the flagged passages makes the system safer but does not bring the answers back. A runtime monitor that checks answers against trusted MITRE data adds a memory with a verifiable safety property and, with a trusted fallback, restores correctness. Detection tells a system what not to trust; getting the answer back needs something it can trust. Future work is a more precise blame step for the monitor, sentence-level checks of answers against trusted statements, stronger adaptive attackers, and larger knowledge bases with more than one generator model.

\section*{Acknowledgment}
This project was supervised by N.~Komninos. The relative-entropy approach was suggested by N.~Komninos, and the runtime-verification monitor by H.~Asad. AI tools were used to help with coding, analysis and drafting; the first author checked all code, numbers and references.

\bibliographystyle{IEEEtran}
\bibliography{refs}

\end{document}